\documentclass[twocolumn,pra,showpacs,preprintnumbers,amsmath,amssymb]{revtex4-1}

\usepackage{graphicx}
\usepackage{dcolumn}
\usepackage{subfigure}
\usepackage{epsfig}
\usepackage{pstricks,pst-grad}
\usepackage{pslatex} 

\def\ket #1{\vert #1\rangle}
\def\bra #1{\langle #1\vert}
\def\ketbra #1#2{\ket{#1}\!\bra{#2}}

\newcommand{\w}{\ensuremath{\mathrm{\omega}}}
\newcommand{\e}{\mathrm{e}}
\renewcommand{\i}{\mathrm{i}}
\newcommand{\SO}{\mathrm{SO}}
\newcommand{\SU}{\mathrm{SU}}
\newcommand{\Zd}{\ensuremath{\mathbb{Z}_d}}
\newcommand{\I}{\mathbb{I}}
\newcommand{\Tr}{\mathrm{Tr}}

\newtheorem{theorem}{Theorem}
\newtheorem{mydef}{Definition}
\newtheorem{lem}[theorem]{Lemma}

\DeclareMathAlphabet{\mathcal}{OMS}{cmsy}{m}{n}

\begin{document}

\title{Noise Thresholds for Higher Dimensional Systems using the Discrete Wigner Function}

\author{Wim van Dam}
\email{vandam@cs.ucsb.edu}
\thanks{}
\affiliation{%
Department of Computer Science, University of California, Santa Barbara, CA 93106, USA\\
Department of Physics, University of California, Santa Barbara, CA 93106, USA
}
\author{Mark Howard}
\email{mhoward@physics.ucsb.edu}
\affiliation{%
Department of Physics, University of California, Santa Barbara, CA 93106, USA
}%
\date{\today}

\begin{abstract}
For a quantum computer acting on $d$-dimensional systems, we analyze the computational power of circuits wherein stabilizer operations are perfect and we allow access to imperfect non-stabilizer states or operations. If the noise rate affecting the non-stabilizer resource is sufficiently high, then these states and operations can become simulable in the sense of the Gottesman-Knill theorem, reducing the overall power of the circuit to no better than classical. In this paper we find the depolarizing noise rate at which this happens, and consequently the most robust non-stabilizer states and non-Clifford gates. In doing so, we make use of the discrete Wigner function and derive facets of the so-called qudit Clifford polytope i.e. the inequalities defining the convex hull of all qudit Clifford gates. Our results for robust states are provably optimal. For robust gates we find a critical noise rate that, as dimension increases, rapidly approaches the the theoretical optimum of 100\%. Some connections with the question of qudit magic state distillation are discussed.
\end{abstract}

\pacs{03.65.Aa, 03.67.-a, 03.67.Pp}
\maketitle

\section{Introduction.}

An interesting question in quantum information theory concerns the types of processes and states that must be present in a quantum circuit that can perform better-than-classical computation. Relevant properties that have been studied include entanglement \cite{JozsaLinden2003} , quantum discord \cite{Eastin:arxiv10} and non-contextuality \cite{Raussendorf:arxiv09}. The Gottesman-Knill theorem \cite{NielsenChuang:2000} tells us that placing a restriction on the states and gates at our disposal can reduce the power of a quantum circuit to no better than classical. Specifically, preparation of $\ket{0}$ states, measurement in the computational basis, feed-forward of measurement results and unitary gates from the Clifford group (including entangling gates) are no more powerful than a classical computer. Collectively, we call these operations stabilizer operations. Clearly, the ability to create non-stabilizer states during the course of a computation is a prerequisite for a quantum-computational speed-up.

The model of quantum computing that we adopt - perfect stabilizer operations plus imperfect non-stabilizer resource - has practical implications for implementing a quantum computer. Buhrman et al. \cite{Buhrmanetal:2006} used these kinds of ideas to show that any qubit-based quantum computer suffering greater than $45\%$ depolarizing noise cannot be more powerful than a classical computer - i.e. they proved an upper bound on the noise threshold for fault-tolerant universal quantum computation (UQC). Plenio and Virmani \cite{Plenio2010} examined the effect of letting noise affect stabilizer operations as well as the non-stabilizer resource, and subsequently showed that the upper bound on the noise rate was on the order of $10\%$ depending on the fault-tolerant scheme one intended to use.

 The lower bound on the noise threshold for fault-tolerant UQC is in the region of (just under) $1\%$ to $5\%$ \cite{AliferisCross07,AliferisPreskill08,Knill05,ReichardtThresh09,Wang10,FujiiYamamoto10} depending on whether these bounds are obtained analytically or via numerical estimates, and depending on particular features of the fault-tolerant scheme (e.g. locality constraints). The noise model employed for such lower bound calculations will also have a significant effect on the values obtained (adversarial noise models are often used in more rigorous estimates, for example). Typically, it seems that the bottleneck lies in implementing the controlled-NOT gate fault-tolerantly. In our model, which assumes perfect stabilizer operations and a non-stabilizer resource subject to depolarizing noise, no such bottleneck exists and it turns out that the lower bound coincides with the upper bound of $45\%$ \cite{ReichardtMagic09,Buhrmanetal:2006}. While the assumption of perfect stabilizer operations may seem naive, it turns out that many proposals for a topological quantum computer using anyons have exactly this property \cite{Georgiev:2006,FreedmanNayakWalker:2006}. In these cases, stabilizer measurements are naturally fault-tolerant, as are Clifford gates, which are implemented by braiding anyons. For such systems, an additional non-stabilizer resource must be introduced in order to achieve UQC, and since it is not topologically protected it is expected to be highly noisy. This was the motivation for Bravyi and Kitaev's investigation of magic state distillation \cite{BravyiKitaev:2005}.

For qubits, the types of resource that provably promote a stabilizer circuit to UQC are fairly well understood. Bravyi and Kitaev \cite{BravyiKitaev:2005}, and subsequently Reichardt \cite{ReichardtMagic09,Reichardt:2005} showed that almost all single qubit non-stabilizer states could perform this task, when sufficiently many copies were input into a magic state distillation routine. The overall effect of this routine is to output an increasingly (with more iterations) pure non-stabilizer state, which in turn allows for the implementation of a unitary gate outside the Clifford group. Campbell and Browne have discussed the structure of distillation protocols \cite{CampbellBrowne:2010} and the impossibility of distilling some non-stabilizer states using known techniques \cite{CampbellBrowne:2009}.  Campbell \cite{Campbell:arxiv10} recently proved the existence of catalysis-like effects for magic state distillation. In a previous work \cite{WvDMH:2009} we showed that, under depolarizing noise, every noisy non-Clifford gate either provably enables UQC or is implementable using Clifford gates only i.e. there is a tight noise threshold. In the qubit case, the states and gates that were most robust to depolarizing noise turned out to be relevant for magic state distillation routines and this is part of the motivation for the current work - little has been done to date on the question of magic state distillation for qudit-based systems. One of our results shows that, even for a hypothetical optimal magic state distillation routine for qudits, the maximum depolarizing noise rate for which a $d$-dimensional state could possibly be distilled is $\frac{d}{d+1}$.

The relevance of the discrete Wigner function (DWF) to our current study arises from a particular definition of DWF due to Galv\~ao \cite{Galvao:2005}, which has the property that the states having a negative DWF are exactly those states that are non-stabilizer \cite{Cormick:2006}. To be more precise, Galv\~ao's construction is based on the DWF definition of Wooters \cite{Wootters1987} and Gibbons \emph{et al.} \cite{Gibbons:2004}, with the additional feature that all possible legitimate DWF in the Gibbons \emph{et al.} construction are considered simultaneously. This property means that Galv\~ao's definition contains redundant information in a tomographic sense (i.e. for description of quantum states) but makes it ideal for our present purposes. While the interpretation of the DWF in terms of discrete points in phase space is a rich subject (many different definitions for DWF exist e.g. that of Gross \cite{Gross:2006} ), we will not discuss this aspect - rather we will use this construction to aid in characterizing non-stabilizer states and non-Clifford operations. It is worth mentioning, however, that two aspects of non-classicality coincide - (i) the existence of negativity in the quasiprobability description of a quantum state and  (ii) the existence of states that are non-classical in the sense of simulability via the Gottesman-Knill theorem. We prove a result conjectured by Wootters (as noted by \cite{Casaccino:2008}) concerning the achievable negativity of quantum states in odd prime dimension. This result involves the explicit construction of so-called phase point operators with a certain spectrum; the more general problem of classifying these phase point operators by spectrum was recently investigated by Appleby \emph{et al.} \cite{appleby:012102}.

We begin by re-stating some relevant facts concerning qudit stabilizer operations and defining our notation. Next, we introduce Galv\~ao's DWF and rewrite it slightly in terms of stabilizer projectors. The final preliminary section discusses how one might go about constructing a polytope whose vertices are Clifford gates. The table at the beginning of section \ref{Results} (Results) represents a summary of the most robust states and gates that we have found. The rest of section \ref{Results} details our method for ascertaining maximally robust states, facets of the Clifford polytope and robust non-Clifford gates respectively.

\section{Preliminaries and Notation.}

\subsection{Qudit Pauli Group and Stabilizer States.}

Throughout, we always assume the dimension $d$ to be a prime number (\cite{Gottesman1999} and \cite{Beigi:arxiv06} are good references for this section).
\begin{equation}
X\ket{j}=\ket{{j+1}\bmod{d}} \quad Z\ket{j}=\w^{j}\ket{j}
\end{equation}
where $\w=\exp(2\pi\i/d)$ is a primitive $d$th root of unity such that $XZ=\w^{-1}ZX$.
Further, we define
\begin{align}
P_{(a|b)}&=X^aZ^b \quad&(d>2)\\
P_{(a|b)}&=i^{ab}\sigma_x^a\sigma_z^b \quad&(d=2)
\end{align}
Using so-called symplectic notation, the general form for multi-particle stabilizer operators
with vectors $x=(x_1,x_2,\dots)$ and $z = (z_1,z_2,\dots)$ with $x_i$, $z_i \in \Zd$ is
\begin{align}
P_{(x|z)}=\left(X^{x_1}\otimes X^{x_2}\dots\right) \left(Z^{z_1}\otimes Z^{z_2}\dots\right).
\end{align}
Two operators commute if and only if the symplectic inner product between their vector descriptions vanishes i.e.\
\begin{align}
&P_{(x|z)} P_{(x^\prime|z^\prime)} = P_{(x^\prime|z^\prime)} P_{(x|z)} \text{ if and only if } \nonumber \\
&\sum_i x_iz^\prime_i-x^\prime_iz_i = x \cdot z^\prime - x^\prime\cdot z = 0\bmod{d}
\end{align}
The Pauli group over $n$ qudits is given by
\begin{align}
\mathcal{G}_n=\left\{\w^c P_{(x|z)} \vert x,z \in \Zd^n, c \in \Zd\right\}
\end{align}
An $n$-qudit stabilizer state $\ket{\psi}$ is the simultaneous eigenvector, with eigenvalue $1$, of a subgroup, $\mathcal{G}_s$, of $\mathcal{G}_n$. The elements of $\mathcal{G}_s$ must be mutually commuting, and $\mathcal{G}_s$ can be generated by $n$ non-trivial elements $\{g_1,g_2,\ldots g_n\}\in \mathcal{G}_n$ i.e.
\begin{align*}
&\mathcal{G}_s=\langle g_1,g_2,\ldots g_n \rangle, \quad |\mathcal{G}_s|=d^n\\
&\ketbra{\psi}{\psi}=\frac{1}{d^n}\sum_{g\in \mathcal{G}_s} g
\end{align*}
where $\langle \cdot \rangle$ denotes a generating set.

The elements of $\mathcal{G}_n$ have eigenvalues from $\left\{\w^k\vert k\in \Zd \right\}$ (replace $\w$ with $-1$ in the rest of this section for the $d=2$ case). Measurement of an arbitrary operator $A$ from $\mathcal{G}_n$ such that the result is $\w^k$ is described by the following projection operator \cite{Gottesman1999}
\begin{align}
\Pi_{[k]}=\frac{1}{d}\left(I+\w^{-k}A+\w^{-2k}A^2+\cdots+ \w^{-(d-1)k}A^{d-1}\right)
\end{align}

We will have occasion to describe both single- and two-qudit projection operators in detail. The $\w^k$ eigenspace of a single-qudit operator $P_{(a|b)}$ corresponds to the projector
\begin{align}
\Pi_{(a|b)[k]}=\frac{1}{d}\left(I+\w^{-k}P_{(a|b)}+\ldots +\w^{-(d-1)k}(P_{(a|b)})^{d-1}\right) \label{Qproj}
\end{align}
which is clearly a qudit stabilizer state (there are a total of $d(d+1)$ distinct single-qudit stabilizer states). Measuring a two-qudit Pauli operator corresponds to projecting with
\begin{align}
\Pi_{(x_1,x_2|z_1,z_2)[k]}=\frac{1}{d}\big(I&+\w^{-k}P_{(x_1,x_2|z_1,z_2)}+\ldots \nonumber \\
\ldots &+\w^{-(d-1)k}(P_{(x_1,x_2|z_1,z_2)})^{d-1}\big)
\end{align}
This eigenspace has dimension $d$ and can be viewed as the codespace of a 2-qudit stabilizer code that encodes one qudit.

Explicit expressions for powers of a Pauli operator can be derived via \cite{Beigi:arxiv06}
\begin{align}
(P_{(x|z)})^m=\w^{\frac{1}{2}m(m-1)x\cdot z}P_{(mx|mz)} \label{Ppower}
\end{align}

\subsection{Discrete Wigner Function.}\label{Discrete Wigner Function}

We use the discrete Wigner function originally defined by Galv\~ao \cite{Galvao:2005}, which is constructed using stabilizer MUB vectors, and which was subsequently shown to have the following property

\begin{theorem}[Cormick \emph{et al.} \cite{Cormick:2006}]
A quantum state (pure or mixed) has a negative Wigner function if and only if the state lies outside the convex hull of stabilizer states.
\end{theorem}

The construction of this DWF relies on associating lines in phase space with vectors (pure stabilizer states) from a set of mutually unbiased bases (MUB). More precisely, for a set of MUB comprising $d+1$ bases, each with $d$ orthonormal states we have a total of $d(d+1)$ states labeled as $\ket{\phi_j^k}$ e.g.
\begin{align*}
\left\{ \ket{\phi_0^1},\ldots\ket{\phi_{d-1}^1},\ket{\phi_0^2}\ldots\ket{\phi_{d-1}^{d+1}}\right\}
\end{align*}

The mutual unbiasedness can then be summarized as
\begin{align*}
\vert \bra{\phi_j^k} \phi_l^m\rangle \vert^2=\frac{1}{d}(1-\delta_{k,m})+\delta_{k,m}\delta_{j,l}
\end{align*}

The Wigner function for a state $\rho$ at a point in phase space $\mathbf{\alpha}$ is given by the expectation value of the so called phase point operator, $A_{\mathbf{\alpha}}$, at that point
\begin{align*}
W_{\mathbf{\alpha}}=\Tr(\rho A_{\mathbf{\alpha}})
\end{align*}
The operator $A_{\mathbf{\alpha}}$ is constructed using the aforementioned MUB vectors $\ket{\phi_j^k}$ (we explain how in detail shortly), and can be viewed as a witness for a state being outside the hull of stabilizer states i.e. a state $\rho$ such that $\Tr(\rho A_{\mathbf{\alpha}})<0$, for any well-defined $A_{\mathbf{\alpha}}$, cannot be decomposed as \begin{align*}
\rho \neq \sum_i q_i \ket{\psi_i^{(s)}}\bra{\psi_i^{(s)}}\quad \left(0\leq q_i \leq 1,~ \sum_i q_i=1\right)
\end{align*}
where the $\ket{\psi_i^{(s)}}$ are $d$-dimensional stabilizer states. The complete set of phase point operators $A_{\mathbf{\alpha}}$ describe bounding inequalities, defining the polytope whose vertices are the $d(d+1)$ stabilizer states that exist in dimension $d$.

There are $d^{(d+1)}$ distinct phase point operators, each of which can be associated with a vector ${u} \in \mathrm{\mathbb{Z}^{(d+1)}_d}$ i.e. $\protect{u=(u_1,u_2,\dots)}$ and $u_i \in \Zd$,
\begin{equation}
A({u})=\frac{1}{d}\left(\sum_{k=1}^{d+1} \ket{\phi_{u_k}^k}\bra{\phi_{u_k}^k}-\I \right) \label{Amub}
\end{equation}

We make the construction of these phase point operators more explicit in the next section. It should be apparent from the above construction that $\Tr(A)=\frac{1}{d}$ always.

\subsubsection{Restatement of DWF.}
If we desire to specify phase point operators $A({u})$ then we must fix a definition for the MUB that we are using. Our definition is not much different to \cite{appleby:012102} except for relabeling and different notation. It is well known that, in prime dimensions $d$, the eigenvectors of the following set of operators constitutes a MUB
\begin{align*}
\{ Z,X,XZ,\ldots XZ^{d-1}\}
\end{align*}
Using the symplectic vector notation for stabilizer operators, we can identify these $d(d+1)$ eigenvectors with projectors indexed by elements $u_i\in \Zd$ e.g
\begin{align}
\{ \Pi_{(0|1)[u_1]},\Pi_{(1|0)[u_2]},\Pi_{(1|1)[u_3]},\ldots \Pi_{(1|d-1)[u_{d+1}]}\}
\end{align}
Now Eq.~\eqref{Amub}
can be rewritten as
\begin{mydef} A phase point operator indexed by ${u}\in \Zd^{(d+1)}$ corresponds to
\begin{align}
A({u})=\frac{1}{d}\left( \Pi_{(0|1)[u_1]}+\sum_{j=2}^{d+1}\Pi_{(1|j-2)[u_j]} -\I \right) \label{Aproj}
\end{align}
where $\Pi_{(a|b)[k]}$ are single-qudit stabilizer projection operators.
\end{mydef}

It is useful to have an expression for the expectation value of $A({u})$ with respect to an arbitrary state $\rho$. First define the Pauli basis coefficients $c_{(a|b)}$ for the state $\rho$
\begin{equation}
c_{(a|b)}=\Tr\big(P_{(a|b)}^\dag \rho\big)
\end{equation}
These $d^2$ coefficients are enough to specify an arbitrary state $\rho$, since the Pauli operators form a complete orthogonal unitary basis.

Using Eq.~\eqref{Qproj} and Eq.~\eqref{Ppower}, one can show that
\begin{align}
\Tr(\rho \Pi_{(a|b)[k]})=\frac{1}{d}\sum_{q=0}^{d-1} \w^{-qk}\w^{-\frac{1}{2}q(q+1)ab}c_{((d-q)a\vert(d-q)b)}
\end{align}
Substituting the projectors that are relevant to $A({u})$ we get
\begin{align}
&\Tr(\rho \Pi_{(0|1)[u_1]})=\frac{1}{d}\sum_{q=0}^{d-1} \w^{-qu_1}c_{(0\vert(d-q)b)} \label{rowproj}\\
&\Tr(\rho \Pi_{(1|0)[u_2]})=\frac{1}{d}\sum_{q=0}^{d-1} \w^{-qu_2}c_{((d-q)a\vert0)} \label{colproj}\\
&\Tr(\rho \Pi_{(1|j-2)[u_j]})=\frac{1}{d}\sum_{q=0}^{d-1} \w^{-qu_j}\w^{-\frac{1}{2}q(q+1)(j-2)}c_{((d-q)\vert(d-q)(j-2))} \label{restproj}
\end{align}
The last equation \eqref{restproj} represents $d-1$ distinct terms, corresponding to the remaining projectors in our definition Eq.~\eqref{Aproj}, indexed by the values $3\leq j \leq d+1$. The expectation value $\Tr(A\rho)$ is proportional to the sum of equations ~\eqref{rowproj} -- \eqref{restproj} and $-\Tr(\rho \mathbb{I})$. Each of Eq.~\eqref{rowproj} -- \eqref{restproj} comprises powers of $\omega$ multiplying Pauli coefficients, $c_{(a|b)}$, and so we can write $\Tr(A\rho)$ in a more intuitive form involving $d \times d$ matrices, i.e.
\begin{align}
\Tr(A\rho)=\frac{1}{d^2}\left(
  \begin{array}{ccc}
    \w^{0} & \w^{u_1} & \hdots \\
    \w^{u_2} & \w^{u_3} & \hdots \\
    \vdots &  & \ddots\\
  \end{array}
\right)\cdot\left(
              \begin{array}{ccc}
                c_{(0|0)} & c_{(0|1)} & \hdots \\
                c_{(1|0)} & c_{(1|1)} &   \\
                \vdots &   & \ddots \\
              \end{array}
            \right)\label{generalform}
\end{align}
where the dot product operation between two matrices $M$ and $N$ is to be interpreted as
\begin{align*}
M \cdot N =\sum_{i,j=1}^{d}M_{i,j}N_{i,j}
\end{align*}
Obviously the left hand matrix of Eq.~\eqref{generalform} is a function of $u \in \Zd^{d+1}$. Moreover, its first row and column are functions of only $u_1$ and $u_2$ respectively, as can be seen by referring to Eq.~\eqref{rowproj} and Eq.~\eqref{colproj}. The remaining $(d-1)^2$ elements of this matrix are functions of $(u_3,u_4,\ldots,u_{d+1})$ and can be calculated using Eq.~\eqref{restproj}.

As a concrete example of the preceding, the expectation value of $A(u_1,u_2,u_3,u_4)$ with respect to a qutrit state $\rho$ (whose Pauli decomposition is $c_{(a|b)}$) is given by
\begin{align*}
\Tr(A\rho)=\frac{1}{3^2}\left(
  \begin{array}{ccc}
    \w^{0} & \w^{u_1} & \w^{2u_1} \\
    \w^{u_2} & \w^{u_3} & \w^{u_4} \\
    \w^{2u_2} & \w^{2u_4+1} & \w^{2u_3+2} \\
  \end{array}
\right)\cdot\left(
              \begin{array}{ccc}
                c_{(0|0)} & c_{(0|1)} & c_{(0|2)} \\
                c_{(1|0)} & c_{(1|1)} & c_{(1|2)} \\
                c_{(2|0)} & c_{(2|1)} & c_{(2|2)} \\
              \end{array}
            \right)
\end{align*}

The equivalent expression for a qubit system $\rho$ and phase point operator $A(u_1,u_2,u_3)$ is
\begin{align*}
\Tr(A\rho)=\frac{1}{2^2}\left(
  \begin{array}{cc}
    (-1)^{0} & (-1)^{u_1} \\
    (-1)^{u_2} & (-1)^{u_3}\\
  \end{array}
\right)\cdot\left(
              \begin{array}{cc}
                c_{(0|0)} & c_{(0|1)} \\
                c_{(1|0)} & c_{(1|1)}\\
              \end{array}
            \right)
\end{align*}
which is probably more readily interpreted when written as
\begin{align}
&\Tr(A\rho)=\frac{1}{4}\left(1+(-1)^{u_1}x+(-1)^{u_2}y+(-1)^{u_3}z\right)\\ &\text{where}\quad  x=\Tr(\sigma_x \rho) \text{ etc.} \nonumber
\end{align}
The interior of the qubit stabilizer octahedron (convex hull of $6$ stabilizer states) is typically parameterized as
\begin{align*}
|x|+|y|+|z|\leq 1
\end{align*}
so we can immediately see that $\Tr(A({u})\rho)\geq 0 \quad ( \forall ~{u}\in \mathbb{Z}_2^3)$ defines the same region.

\subsubsection{Robustness of States to Depolarizing Noise}
If a state $\rho$ is outside the convex hull of stabilizer states then we must have $\Tr(\rho A({u}))<0$ for at least one of the phase point operators. We define the negativity of a state as
\begin{mydef}{Negativity of a state, $\rho$ is denoted $|N(\rho)|$:}
\begin{align}
|N(\rho)|= \begin{cases}&\left|~\underset{{u}\in \Zd^{d+1}}{\min}\left[ \Tr(\rho A({u}))\right]\right|\\
&0 \iff \Tr(\rho A({u}))\geq 0, \forall~{u}\in \mathbb{Z}_d^{d+1}
\end{cases}
\end{align}
\end{mydef}

\begin{mydef}{Robustness to depolarizing noise of a state $\rho$ is denoted $p^\star(\rho)$ :}
\begin{align*}
&p^\star(\rho)= \min(p) \text{ such that }\\
&(1-p)\rho + p\frac{\I}{d} = \sum_i q_i \ket{\psi_i^{(s)}}\bra{\psi_i^{(s)}}\quad \left(0\leq q_i \leq 1,~ \sum_i q_i=1\right)
\end{align*}
where the $\ket{\psi_i^{(s)}}$ are $d$-dimensional stabilizer states.
\end{mydef}

\begin{lem} Negativity of a state $\rho$ and its robustness to noise are related by
\begin{align}
p^\star(\rho) = 1- \frac{1}{d^2|N(\rho)|+1}\nonumber
\end{align}
\end{lem}

First note that since the trace of a phase point operator is always $\frac{1}{d}$, then the maximally mixed state $\frac{\I}{d}$ satisfies
\begin{align*}
\Tr\left(A({u}) \frac{\I}{d}\right)=\frac{1}{d^2},\quad \forall~{u}\in \mathbb{Z}_d^{d+1}
\end{align*}
Let $A$ be the phase point operator that minimizes $\Tr(A\rho)$ for a particular state $\rho$, then
\begin{align}
&\Tr\left(   A \left[ (1-p)\rho +p\frac{\I}{d} \right]  \right)\geq 0\\
&\iff -(1-p)|N(\rho)|+\frac{p}{d^2}\geq 0\\
&\iff p\geq 1- \frac{1}{d^2|N(\rho)|+1}&\\
&\Rightarrow  p^\star(\rho) = 1- \frac{1}{d^2|N(\rho)|+1}&
\end{align}
Later we will use this expression to prove the existence of states that can survive depolarizing rates of up to $p=\frac{d}{d+1}$ before becoming stabilizer states.

\subsection{Noisy Quantum Operations and the Clifford Polytope}
We saw in section \ref{Discrete Wigner Function} how the set of states inside the convex hull of stabilizer states can be described by bounding inequalities, where each such inequality corresponds to a Hermitean operator, $A$. We have previously argued that access to states from inside this region does not improve the computational power of a quantum computer that can only implement stabilizer operations (albeit perfectly). For the same reason, access to operations inside the convex hull of Clifford gates is equally unhelpful. The structure of the qudit Clifford polytope is of inherent interest, but we are especially interested in what it tells us about operations that are highly robust to noise. We denote a noisy (depolarized) version of non-Clifford gate, $U$, as $\mathcal{E}_U$ i.e.
\begin{align}
&\mathcal{E}_U:\rho \rightarrow \rho^\prime \nonumber\\
&\text{where } \rho^\prime = (1-p)U\rho U^\dag +p \frac{\I}{d}
\end{align}
Similarly to how we defined states' robustness to noise in the previous section, we define the robustness of a non-Clifford gate $U$ as the minimum noise rate that takes it inside the Clifford polytope
\begin{mydef}{Robustness to depolarizing noise of an operation $U$ is denoted $p^\star(U)$ :}
\begin{align}
&p^\star(U)= \min(p) \text{ such that } \nonumber \\
&(1-p)U\rho U^\dag +p \frac{\I}{d} = \sum_i q_i C_i \rho C_i^\dag \quad \left(0\leq q_i \leq 1,~ \sum_i q_i=1\right) \nonumber
\end{align}
where the $C_i$ are Clifford gates. In other words, $p^\star(U)$ is the noise rate at which $\mathcal{E}_U$ enters the Clifford polytope.
\end{mydef}

Finding the bounding inequalities for the convex hull of a set of vertices is known as the halfspace enumeration problem. The reverse question, finding the vertices given a set of inequalities is known as vertex enumeration. Both the halfspace and vertex enumeration problems get rapidly more difficult as the size of the problem increases e.g.\ computing the convex hull of $n$ points in $d$-dimensional space requires $\mathcal{O}(n^{\lfloor \frac{d}{2} \rfloor})$ time \cite{Chazelle1991}. Various software implementations exist to solve these kinds of problems (e.g \cite{Avis2000Revised,Fukuda1996}). In order to find the convex hull of Clifford gates using this software, one must first describe the gates using an appropriate real, linear representation. For example, it is possible to represent each Clifford gate $C \in \SU(d)$ as an orthogonal rotation matrix $R_C \in \SO(d^2-1)$ \cite{Bagan:2003}. Alternatively, one can represent the Choi-Jamio\l kowski state corresponding to each Clifford gate by using the well known construction for power-of-prime MUBs (e.g. \cite{Bandyopadhyay2008}).

Using one of the aforementioned software packages, Buhrman et al. \cite{Buhrmanetal:2006} derived a complete list of facets (non-redundant halfspace inequalities) for the polytope whose vertices are the $24$ single-qubit Clifford gates - the so-called Clifford polytope. Doing so allowed them to find the non-Clifford gate that was most resistant to noise. One of the main goals of this work is to examine the structure of the qudit version of Buhrman et al.'s Clifford polytope \cite{Buhrmanetal:2006}, i.e. the structure in $(d^2-1)^2$-dimensional space whose vertices comprise the set of $d^3(d^2-1)$ single-particle Clifford operations on $d$-dimensional systems. Unfortunately, because of the aforementioned computational complexity of the halfspace enumeration problem we were unable to complete the calculation that would provide all the facets of the qudit Clifford polytope for any $d>2$. However, building on our previous work, we predicted that the facets would have a certain structure and could be logically deduced. Checking that a conjectured facet is indeed a true facet is not computationally difficult, and so we were able to verify our conjectured facets, and consequently find a large set of distinct facets for $d=3,5,7$. There is precedence for deriving facets of a polytope in quantum information theory, typically in the setting of non-locality, whereby the vertices correspond to measurement outcomes of different parties and facets are tight Bell inequalities (see e.g. \cite{Masanes2003}).

\subsubsection{Checking Conjectured Facets}\label{Checking Conjectured Facets}

A polytope $\mathcal{P}$, in $D$ dimensions, can be uniquely characterized by a finite set of bounding inequalities called facets, $\{{F}_i, f_i\}$, where $F_i \in \mathbb{R}^D$ and $f_i \in \mathbb{R}$. We can test membership in $\mathcal{P}$ of some vector ${X}$ by taking the dot product with these facet-defining vectors $F_i$
\begin{align*}
{X} \in {P} \iff {X} \cdot {F}_i \leq f_i \quad \forall i
\end{align*}
Let us denote the vertices of this polytope as ${V}_j$ - again there are a finite number of these.

The Clifford polytopes that we study will live in a space of dimension $D=(d^2-1)^2$, where $d$ is the dimension of the qudit system that is under consideration. If we construct a conjectured facet $\{G,g\}$, then for it to be a true facet (see e.g. \cite{Masanes2003}) it must satisfy
\begin{align*}
\textbf{Condition 1:  } \qquad \forall~ {V}_j,\quad {V}_j\cdot {G} \leq g
\end{align*}
and, defining
\begin{align*}
\{\widetilde{{V}_j}\}= \{{V}_j\vert {V}_j\cdot {G} = g\}
\end{align*}
we must also have
\begin{align*}
\textbf{Condition 2: }\quad \operatorname{rank}\left(
                     \begin{array}{ccc}
                       \leftarrow  & \widetilde{{V}_1} & \rightarrow \\
                       \leftarrow  & \widetilde{{V}_2} & \rightarrow  \\
                        & \vdots &   \\
                     \end{array}
                   \right) = D
\end{align*}
i.e. we must have $D$ linearly independent vertices contained in ${G}$. Later we will rewrite this second condition simply as $\operatorname{rank}\{\widetilde{{V}_j}\}=D$.

\section{Results}\label{Results}

\subsection{Summary of Results}
Table I summarizes the robustness of non-stabilizer states and non-Clifford gates to depolarizing noise. The rest of this results section addresses how these values were determined. We first deal with the left half of the table (robust states) before moving on to the right half (robust gates).

\renewcommand\arraystretch{2}
\begin{table}
\begin{tabular*}{0.48\textwidth}{@{\extracolsep{\fill}}|c|c|c||c|c|}
  \hline
        & $\mathbf{p^\star(\rho)}$ & \textbf{Optimal?} & $\mathbf{p^\star(U)}$ & \textbf{Optimal?}  \\ \hline
  $\mathbf{d=2}$ & $1-\tfrac{1}{\sqrt{3}}\approx 42\%$ & $\checkmark$ & $1-\frac{1}{2\sqrt{2}-1}\approx 45.3\%$ & $\checkmark$  \\ \hline
  $\mathbf{d=3}$ & $\frac{3}{4}=75\%$ &  $\checkmark$ & $\approx 78.6\%$ & ?  \\ \hline
  $\mathbf{d=5}$ & $\frac{5}{6}\approx 83\%$ &  $\checkmark$ & $1-\frac{1}{21}\approx 95.2\%$ & ?  \\ \hline
  $\mathbf{d=7}$ & $\frac{7}{8}=87.5\%$ &  $\checkmark$ & $\approx 97.6\%$ & ? \\ \hline
  $\mathbf{d}$ & $\frac{d}{d+1}$ &  $\checkmark$ &  &   \\
  \hline
\end{tabular*}\label{resultstable}
\caption{Summary of results: The maximal robustness to depolarizing noise of non-stabilizer states $p^\star(\rho)$, and robustness of non-Clifford gates $p^\star(U)$. }
\end{table}

\subsection{Robust Qudit States.}

This subsection comprises the proof of the following theorem and a discussion of some of its implications.

\begin{theorem}\label{RobustState}
For all odd prime dimensions, $d$, there is a family of pure states, $\{\ket{\nu}\}$, which require a depolarizing rate of $p=\frac{d}{d+1}$ in
\begin{align*}
(1-p)\ket{\nu}\bra{\nu}+p\frac{\I}{d}
\end{align*}
to become a stabilizer state. Furthermore, these states are maximally robust amongst all states in dimension $d$.
\end{theorem}

We will find a family of states for all prime dimension $d$ that are maximally robust to depolarizing noise. Using the previously-derived relationship between robustness ($p^\star$) and negativity ($|N|$),
\begin{align}
p^\star(\rho) = \frac{1}{d^2|N(\rho)|+1}
\end{align}
this amounts to finding states that have maximal negativity in terms of their discrete Wigner function. It was conjectured by Wootters that the maximal negativity achievable by quantum states in odd prime dimension is $|N(\rho)|=\frac{1}{d}$.
Recall that the Wigner function of a state is its expectation value with respect to a phase point operator $A({u})$. We can use this to show that negativity is bounded above by $|N(\rho)|\leq\frac{1}{d}$
\begin{align}
W_{{u}}(\rho)&=\Tr(A({u})\rho )\\
&=\frac{1}{d}\left(\Tr(\rho \Pi_{(0|1)[u_1]})+\sum_{j=2}^{d+1}\Tr(\rho\Pi_{(1|j-2)[u_j]})-\Tr(\rho\I) \right)\\
&=\frac{1}{d}\left(\sum_{i=1}^{d+1} q_i -1\right)\quad (0\leq q_i \leq 1)\\
&\Rightarrow  \underset{{u}, ~ \rho}{\min}~ \left[W_{{u}}(\rho)\right]\geq-\frac{1}{d}\\
& \Rightarrow |N(\rho)|\leq\frac{1}{d}
\end{align}
Next, we show that this limit is achievable in all odd prime dimensions.

Casaccino, Galv\~ao and Severini \cite{Casaccino:2008} proved that the quantity
\begin{align}
\underset{\rho}{\min}\left[ \Tr(A \rho) \right]
\end{align}
is minimized by setting
\begin{align}
&\rho = \ket{\nu_1}\bra{\nu_1} \text{ where } A=\sum_{k=1}^{d} \lambda_k \ket{\nu_k}\bra{\nu_k} \\
&\text{ and } \lambda_1 \leq \lambda_2 \ldots \leq \lambda_d
\end{align}
In other words $\Tr(A \rho)$ is minimized by finding the minimal eigenvalue of $A$ and using the corresponding normalized eigenvector for $\rho$ (this is actually a common technique to find a state that minimizes or maximizes an operator). To complete the proof that $|N(\rho)|=\frac{1}{d}$ is achievable, we need to show that there is always a phase point operator, in odd prime dimension $d$, with at least one eigenvalue $\lambda_1=-\frac{1}{d}$.

Bandyopadhyay \emph{et al.} \cite{Bandyopadhyay2008} showed that the eigenvector (with eigenvalue $\w^k$) of $X Z^b$ is given by
\begin{align}
&\ket{\psi_k^b}=\frac{1}{\sqrt{d}}\sum_{m=0}^{d-1} (\w^k)^{d-m}(\w^{-b})^{s_m} \ket{m}\\
&\text{where } s_m=\sum_{q=m}^{d-1} q
\end{align}

Since
\begin{align}
\ket{\psi_0^b}\bra{\psi_0^b}&=\Pi_{(1|b)[0]}\\
\text{and }\qquad \ket{k}\bra{k}&=\Pi_{(0|1)[k]}
\end{align}
we have the following matrix form for the phase point operator corresponding to ${u}^\star=(\frac{d+1}{2},0,\ldots 0)$
\begin{align*}
A({u}^\star)&=\frac{1}{d}\left(\ket{\tfrac{d+1}{2}}\bra{\tfrac{d+1}{2}}+\frac{1}{d}\sum_{b=0}^{d-1} \ket{\psi_0^b}\bra{\psi_0^b}-\I\right)\\
&=\frac{1}{d}\left(\ket{\tfrac{d+1}{2}}\bra{\tfrac{d+1}{2}}+\frac{1}{d}\sum_{b,m,n=0}^{d-1} \omega^{b(s_{n}-s_m)}\ket{m}\bra{n}-\I\right)\\
&\text{where } s_m=\sum_{q=m}^{d-1} q \quad \text{and} \quad  s_{n}=\sum_{q=n}^{d-1} q \qquad
\end{align*}

For a moment let us concentrate on the matrix
\begin{align}
\frac{1}{d}\sum_{b,m,n=0}^{d-1} \omega^{b(s_{n}-s_m)}\ket{m}\bra{n}=\sum_{m,n=0}^{d-1} \left(\frac{1}{d} \sum_{b=0}^{d-1} \omega^{b(s_{n}-s_m)}\ket{m}\bra{n}\right)
\end{align}
Use the facts that
\begin{align*}
&\frac{1}{d} \sum_{b=0}^{d-1} \omega^{b(s_{n}-s_m)}=\begin{cases} 0: \quad s_{n} \neq s_m \\
1: \quad s_{n} = s_m \end{cases}\\
\text {and } &s_m=\sum_{q=m}^{d-1}q=\frac{1}{2}(m-m^2) \mod d\\
\end{align*}
to show that non-zero coefficients occur only when $\protect{ m(m-1)=n(n-1)}$ i.e. when
\begin{align*}
 &n=m \mod d\\
\text{or }  &n=1-m \mod d
\end{align*}

Taking all of the above into consideration, the matrix $\frac{1}{d}\sum_{b,m,n=0}^{d-1} \omega^{b(s_{n}-s_m)}\ket{m}\bra{n}$
takes a particularly simple form -- it is block diagonal with a $2\times 2$ submatrix of ones and another submatrix with ones on the diagonal and antidiagonal. For example for $d=7$ we have
 \begin{eqnarray*}
\frac{1}{d}\sum_{b,m,n=0}^{d-1} \omega^{b(s_{n}-s_m)}\ket{m}\bra{n}= \renewcommand{\arraystretch}{1}
 \left( \begin{array}{cc|ccccc}
   \renewcommand{\arraystretch}{1.2}1 & 1 & 0 & 0 & 0 & 0 & 0 \\
   1 & 1 & 0 & 0 & 0 & 0 & 0 \\ \hline
   0 & 0 & 1 & 0 & 0 & 0 & 1 \\
   0 & 0 & 0 & 1 & 0 & 1 & 0 \\
    0 & 0 & 0 & 0 & 1 & 0 & 0 \\
    0 & 0 & 0 & 1 & 0 & 1 & 0 \\
    0 & 0 & 1 & 0 & 0 & 0 & 1
 \end{array} \right)\quad (d=7)
 \end{eqnarray*}
Note that the diagonal and anti-diagonal of the second submatrix intersect at $m=\frac{d+1}{2}$.

Returning to the phase point operator $A({u}^\star)$ we see that subtracting the identity and adding in $\ket{\tfrac{d+1}{2}}\bra{\tfrac{d+1}{2}}$ gives us a matrix of the form
\begin{eqnarray*}
A({u}^\star)=\renewcommand{\arraystretch}{1}
\frac{1}{7} \left( \begin{array}{cc|ccccc}
   \renewcommand{\arraystretch}{1.2}0 & 1 & 0 & 0 & 0 & 0 & 0 \\
   1 & 0 & 0 & 0 & 0 & 0 & 0 \\ \hline
   0 & 0 & 0 & 0 & 0 & 0 & 1 \\
   0 & 0 & 0 & 0 & 0 & 1 & 0 \\
    0 & 0 & 0 & 0 & 1 & 0 & 0 \\
    0 & 0 & 0 & 1 & 0 & 0 & 0 \\
    0 & 0 & 1 & 0 & 0 & 0 & 0
 \end{array} \right)\quad (d=7)
 \end{eqnarray*}
i.e.\ it contains two counter-identity matrices of sizes $2\times 2$ and $(d-2) \times (d-2)$ respectively. It is known that $\lfloor \frac{n}{2} \rfloor$ of the eigenvalues of an $n \times n$ counter-identity matrix are $-1$ (and the rest are $1$). In our case this means that $\frac{d-3}{2}+1=\frac{d-1}{2}$ eigenvalues of $A({u}^\star)$ are $-\frac{1}{d}$. Each corresponding eigenvector $\ket{\nu_k}$ (properly normalized) is a state that is maximally robust to depolarizing noise.
\begin{align}
&A({u}^\star)=\sum_{k=1}^{d} \lambda_k \ket{\nu_k}\bra{\nu_k} \quad (\lambda_1 \leq \lambda_2 \ldots \leq \lambda_d)\\
& \text{ So, for } ~k\in\left\{1,2\ldots\frac{d-1}{2}\right\} \text{ we have }\nonumber \\
&\Tr(A({u}^\star) \ket{\nu_k}\bra{\nu_k})=-\frac{1}{d} \\
&\Rightarrow  p^\star(\ket{\nu_k}\bra{\nu_k}) = 1-\frac{1}{d^2\left(\frac{1}{d}\right)+1} = \frac{d}{d+1}
\end{align}

It was noted in \cite{appleby:012102} that each $A$ is part of a subset of phase point operators of size $d^2$, all of which have the same spectrum. We can generate the remaining $d^2-1$ remaining phase point operators related to $A({u}^\star)$ by applying each of the $d^2-1$ non-trivial Pauli operators to it. Taking the $\frac{d-1}{2}$ eigenvectors, with eigenvalue $-\frac{1}{d}$, of all phase point operators related to $A({u}^\star)$ gives us a total of $\frac{d^2(d-1)}{2}$ maximally robust states.

Let us denote by $\ket{\nu_1}$ the following eigenvector of $A({u}^\star)$
\begin{align}
\ket{\nu_1}=\frac{\ket{0}-\ket{1}}{\sqrt{2}} \qquad \left(\lambda_1=-\frac{1}{d}\right)
\end{align}
We show in the appendix that this vector is an eigenvector of a qudit Clifford operation. The so-called $\ket{T}$-type magic states for qubits (which are known to be distillable for less than about $35\%$ depolarizing noise) are qubit states that are maximally robust to depolarizing noise, and are also eigenvectors of Clifford operations. It is an interesting open question whether the maximally robust qudit states, $\ket{\nu_1}$, that we have just described are distillable by some qudit magic state distillation routine.

\subsection{Method for Deriving Facets of the Clifford Polytope.}\label{Method for Deriving Facets of the Clifford Polytope.}

We describe a method of deducing facets of the Clifford polytope by using a simple argument:
\begin{quotation}If some two-qudit state $\tau$ is acted on by stabilizer operations only (actually this process is the decoding for a two-qudit stabilizer code), and the output of this process is a single-qudit non-stabilizer state $\rho$, then the original two-qudit state $\tau$ cannot have been a stabilizer state.
\end{quotation}
Note that the converse is not necessarily true.

\begin{mydef} A Choi-Jamio\l kowski state corresponding to a unitary operation $U$ on a qudit state is denoted
\begin{align}
\ket{J_U}=(\I\otimes U) \sum_{j=0}^{d-1} \frac{\ket{jj}}{\sqrt{d}}
\end{align}
\end{mydef}

\begin{mydef} A weight-2 two-qudit $\tau$ state is one for which
\begin{align}
&c_{(x_1,x_2|z_1,z_2)}=0 \text{ if } x_1=z_1=0 \text { or }  x_2=z_2=0 \nonumber \\
&\text{where } c_{(x_1,x_2|z_1,z_2)}=\text{\emph{Tr}}(P_{(x_1,x_2|z_1,z_2)}^\dag \tau) \nonumber
\end{align}
i.e. states $\tau$ are those for which a local Pauli measurement must have zero expectation value.
\end{mydef}

Since the identity coefficient $c_{(0,0|0,0)}$ is identically $1$, we have potentially $(d^2-1)^2$ non zero coefficients. The corresponding weight-two Pauli operators form a unitary orthogonal basis for the space of Choi-Jamio\l kowski states.

The set of states $\{ \ket{J_{C_i}} \}$, where $\{ C_1,C_2 \ldots C_{d^3(d^2-1)}\}$ are all the Clifford gates for a single qudit, form the vertices of the Clifford polytope. The noisy operation $\mathcal{E}_U$ (that we want to check for membership of the Clifford polytope) will be encoded in the weight-two operator $\tau$ via
\begin{align}
\tau=(\mathcal{I}\otimes \mathcal{E}_U) \left(\frac{1}{d}\sum_{j,k=0}^{d-1} \ket{jj}\bra{kk}\right)=\frac{1}{d^2}c_{(x_1,x_2|z_1,z_2)}P_{(x_1,x_2|z_1,z_2)}
\end{align}

If we somehow had access to a complete facet description, $\mathcal{F}$, of the Clifford polytope (where the individual facets are written as Hermitean operators, $F$, in $d \times d$-dimensional Hilbert space) then we could immediately test whether or not $\mathcal{E}_U$ is a Clifford operation via
\begin{align}
\exists F \text{ such that } \Tr(F \tau)<0 \iff \mathcal{E}_U \text{ is non-Clifford}
\end{align}
In the absence of a complete facet description, we instead create operators for which the forward implication is necessarily true. In turn, we show that these operators are indeed facets - although it is possible our method does not enable construction of every $F \in \mathcal{F}$.

\begin{mydef} A Clifford witness $W$ is a Hermitean operator satisfying
\begin{align}
&\Tr(W \tau)<0 \Rightarrow \mathcal{E}_U \text{ is non-Clifford}\\
&\text{where }\tau=(\mathcal{I}\otimes \mathcal{E}_U) \left(\frac{1}{d}\sum_{j,k=0}^{d-1} \ket{jj}\bra{kk}\right) \nonumber
\end{align}
\end{mydef}

The general idea for our method of constructing Clifford witnesses is as follows
\begin{enumerate}
\item Project arbitrary weight-2 state $\tau$ into codespace of two-qudit stabilizer code. This produces an encoded state $\overline{\rho}$.
\item Decode $\overline{\rho} \rightarrow \rho $ using stabilizer operations only.
\item Derive conditions on coefficients $c_{(x_1,x_2|z_1,z_2)}$ of $\tau$ such that $\Tr(\rho A({u})<0$.
\end{enumerate}

Since decoding a stabilizer code uses only stabilizer operations, we know that any $\tau$ satisfying these conditions on $c_{(x_1,x_2|z_1,z_2)}$ must have been non-stabilizer to begin with. Each of these conditions on $c_{(x_1,x_2|z_1,z_2)}$ corresponds to a hyperplane in $(d^2-1)^2$-dimensional space - where one side of the hyperplane contains only $\tau$ corresponding to non-Clifford operations.  What we want is a facet -- a hyperplane that forms part of the boundary of the convex hull of Clifford operations - so we need to check that conditions $1$ and $2$ described in Section \ref{Checking Conjectured Facets} are satisfied.

There are many different codes we could choose so for simplicity we concentrate on codes defined by the projector
\begin{align}
\Pi_{(0,0|1,1)[0]}&=  \sum_{j=0}^{d-1} \ket{\overline{j}}\bra{\overline{j}}\\
\text{where }\quad \ket{\overline{j}}\bra{\overline{j}} &= \frac{1}{d^2}\langle P_{(0,0|1,1)}, \w^jP_{(0,0|0,1)} \rangle \label{paritycode}
\end{align}
which projects onto the parity-zero subspace of the two-qudit Hilbert space. These will eventually be used to derive what we have called ``$B$-type" facets in previous work. We also use the codes
\begin{align}
\Pi_{(0,0|0,1)[0]}&=  \sum_{j=0}^{d-1} \ket{\overline{j}}\bra{\overline{j}}\\
\text{where }\quad \ket{\overline{j}}\bra{\overline{j}} &= \frac{1}{d^2}\langle P_{(0,0|0,1)}, \w^jP_{(0,0|1,0)} \rangle\\
\Pi_{(0,0|1,0)[0]}&=  \sum_{j=0}^{d-1} \ket{\overline{j}}\bra{\overline{j}}\\
\text{where }\quad \ket{\overline{j}}\bra{\overline{j}} &= \frac{1}{d^2}\langle P_{(0,0|1,0)}, \w^jP_{(0,0|0,1)} \rangle
\end{align}
to derive different types of facets, which were called ``$A$-type" facets and ``$A^T$-type" facets in previous work \cite{WvDMH:2009}.

Conjugating any facet by local Clifford operations must produce another (not necessarily distinct) facet, so each facet, $F$, that we derive can immediately be used to generate a family of facets
\begin{align}
&\mathcal{F}=\{F^\prime \vert F^\prime=(C_i \otimes C_j) F (C_i \otimes C_j)^\dag\} \\ &\forall~i,j\in\{1,2,\ldots,d^3(d^2-1)\}\nonumber
\end{align}
all of which will have the same spectrum (by construction).
\begin{theorem}\label{QubitComplete}
The preceding method for deriving facets of the Clifford polytope gives a complete description (i.e. all possible facets), for the qubit ($d=2$) case. Furthermore, the optimal non-Clifford gate and the threshold depolarizing rate are given by an eigenvector and eigenvalue of the Hermitean Clifford witnesses $W_B$ respectively.
\end{theorem}
The proof is contained in the following example and the subsequent section on finding robust operations.

\subsubsection{Example: Deriving Qubit (d=2) Clifford Polytope Facets}\label{Example: Deriving Qubit (d=2) Clifford Polytope Facets}

Define
\begin{align}
\overline{\rho}=\frac{\Pi_{(0,0|1,1)[0]} \tau \Pi_{(0,0|1,1)[0]}}{\Tr\left(\Pi_{(0,0|1,1)[0]} \tau \Pi_{(0,0|1,1)[0]}\right)}
\end{align}
and decoding this to a single qubit produces

\begin{align}
&\overline{\rho}\underset{\text{dec}}{\rightarrow}\rho=\sum_{m,n=0}^{d-1} \bra{\overline{m}}\overline{\rho}\ket{\overline{n}}\ket{m}\bra{n} = \nonumber\\
&\sum_{k\in\mathbb{Z}_2} \left(
                                                        \begin{smallmatrix}
                                                          c_{(0,0|k,k)} & (-1)^kc_{(1,1|k,k)}+(-i)^kc_{(1,1|k+1,k)} \\
                                                          (-1)^kc_{(1,1|k,k)}+(i)^kc_{(1,1|k+1,k)}  & c_{(0,0|k,k)} \\
                                                        \end{smallmatrix}
                                                      \right)
\end{align}

Calculating the expectation value for a phase point operator with respect to the normalized single-qubit $\rho$ gives e.g.
\begin{align}
&\text{For } ~{u}=(0,0,0):\quad A({u})=\left(
           \begin{array}{cc}
             \frac{1}{2} & \frac{1}{4}-\frac{i}{4} \\
             \frac{1}{4}+\frac{i}{4}  & 0 \\
           \end{array}
         \right)\\
&\text{and }~ \Tr(\rho A({u}))=\frac{\sum_{k\in\mathbb{Z}_2}\left(c_{(0,0|k,k)}+c_{(1,1|k+1,k)}+(-1)^kc_{(1,1|1,k)}\right)}{\sum_{k\in\mathbb{Z}_2}c_{(0,0|k,k)}}
\end{align}

Because we are only interested in the sign of $\Tr(\rho A({u}))$ the denominator is unimportant, and so any state $\tau$ that satisfies
\begin{align}
\sum_{k\in\mathbb{Z}_2}\left(c_{(0,0|k,k)}+c_{(1,1|k+1,k)}+(-1)^kc_{(1,1|1,k)}\right)<0
\end{align}
must not correspond to a Clifford operation. Since $c_{(x_1,x_2|z_1,z_2)}=\Tr(P_{(x_1,x_2|z_1,z_2)}^\dag \tau)$ (by definition) the condition on $\tau$ can be expressed as the expectation of an observable i.e. $\Tr(W_B \tau)<0$ implies a non-Clifford operation, where
\begin{align}
&W_B=\sum_{k\in\mathbb{Z}_2}\left(P_{(0,0|k,k)}^\dag+P_{(1,1|k+1,k)}^\dag+(-1)^kP_{(1,1|1,k)}^\dag\right)\\
&\Rightarrow \qquad\quad W_B=\left(
    \begin{array}{cccc}
      2 & 0 & 0 & 2-2i \\
      0 & 0 & 0 & 0 \\
      0 & 0 & 0 & 0 \\
     2+2i & 0 & 0 & 2 \\
    \end{array}
  \right)\label{WBqubit}
\end{align}

We omit the details, but the same kind of analysis, using the encoding
\begin{align}
\overline{\rho}\propto \Pi_{(0,0|0,1)[0]} \tau \Pi_{(0,0|0,1)[0]}
\end{align}
and the same phase point operator $A(0,0,0)$, shows us that any $\tau$ satisfying
\begin{align}
c_{(0,0|0,0)}+c_{(0,0|1,1)}+c_{(1,0|0,1)}+c_{(1,0|1,1)}<0
\end{align}
must represent a non-Clifford operation. This inequality implies a witness of the form
\begin{align}
W_A=\left(
    \begin{array}{cccc}
      2 & 0 & 1-i & 0 \\
      0 & 0 & 0 & -1+i \\
      1+i & 0 & 0 & 0 \\
     0 & -1-i & 0 & 2 \\
    \end{array}
  \right)
\end{align}
Finally, using the closely related encoding
\begin{align}
\overline{\rho}\propto \Pi_{(0,0|1,0)[0]} \tau \Pi_{(0,0|1,0)[0]}
\end{align}
we find that any $\tau$ satisfying
\begin{align}
c_{(0,0|0,0)}+c_{(0,0|1,1)}+c_{(0,1|1,0)}+c_{(0,1|1,1)}<0
\end{align}
must represent a non-Clifford operation. The witness is
\begin{align}
W_{A^{T}}=\left(
    \begin{array}{cccc}
      2 & 1-i & 0 & 0 \\
      1+i & 0 & 0 & 0 \\
      0 & 0 & 0 & -1+i \\
     0 & 0 & -1-i & 2 \\
    \end{array}
  \right)
\end{align}

Next, we seek to ascertain whether these witnesses are tight against the convex hull of Clifford operations. The following facts, which are easily checked, ensure that the witnesses are tight - i.e. they are facets of the Clifford polytope.
\begin{align}
&\text{For all } F\in\{A,A^T,B\} \nonumber \\
&\bra{J_{C_i}} W_F \ket{J_{C_i}} \geq 0 \quad \forall~ i\in\{1,2,..,24\} \label{cond1} \\
&\text{and, defining the subset } \nonumber\\
&\left\{\ket{\widetilde{J_{C_i}}}\right\}_F=
\bigl\{\ket{J_{C_i}} ~ \bigm\vert ~ \bra{J_{C_i}} W_F \ket{J_{C_i}} = 0\bigr\} \nonumber\\
&\operatorname{rank}\left\{\ket{\widetilde{J_{C_i}}} \right\}_F = 9 = (d^2-1)^2 \label{cond2}
\end{align}
Equations \eqref{cond1} and \eqref{cond2} correspond to conditions $1$ and $2$ described in Section\ref{Checking Conjectured Facets}.
It should also be noted that
\begin{align}
&\left\vert \left\{\ket{\widetilde{J_{C_i}}}\right\}_A \right\vert=\left\vert \left\{\ket{\widetilde{J_{C_i}}} \right\}_{A^T} \right\vert=12 \nonumber,\\
&\left\vert \left\{\ket{\widetilde{J_{C_i}}} \right\}_B \right\vert=14
\end{align}
i.e. $A^{(T)}$-type facets contain $12$ Clifford vertices each, while $B$-type facets contain $14$.

As a final step, we generate sets of facets from these canonical representatives
\begin{align}
\mathcal{W}_F=\{W_F^\prime \vert W_F^\prime=(C_i \otimes C_j) W_F (C_i \otimes C_j)^\dag\}\\ \forall~i,j\in\{1,2,\ldots,24\},\quad F\in\{A,A^T,B\} \nonumber
\end{align}
resulting in the set of facets
\begin{align}
\mathcal{S}=\mathcal{W}_A \cup \mathcal{W}_{A^T} \cup \mathcal{W}_{B}
\end{align}
It is straightforward to verify that $\mathcal{S}$ contains $|\mathcal{S}|=120$ distinct facets, and, using software for vertex enumeration, that the polytope represented by $\mathcal{S}$ has all 24 Clifford operations as its vertices - i.e $\mathcal{S}$ is exactly the Clifford polytope arrived at by Buhrman et al. \cite{Buhrmanetal:2006}.

\subsection{Robust Qudit Operations.}
If we have a facet of the Clifford polytope, how do we find the operation that is maximally robust to depolarizing noise before it gets pushed inside the polytope? Recall that a two-qudit state $\tau$ represents a non-Clifford operation $\mathcal{E}_U$ if
\begin{align}
&\Tr(\tau W)<0 \quad\text{where} \\
&\tau=(\mathcal{I}\otimes \mathcal{E}_U) \left(\frac{1}{d}\sum_{j,k=0}^{d-1} \ket{jj}\bra{kk}\right)=(1-p)\ketbra{J_U}{J_U} +p\frac{\I}{d^2}
\end{align}
so clearly $\langle J_U \vert W \vert J_U \rangle <0 $, and the more negative this quantity, the greater the depolarizing rate required to make $\Tr(\tau W)=0$.

Let $U$ be the unitary that minimizes $\langle J_U \vert W \vert J_U \rangle$, and re-scale $W$ to make it have unit trace, then
\begin{align}
&\Tr\left(   W \left[ (1-p)\ketbra{J_U}{J_U} +p\frac{\I}{d^2} \right]  \right)\geq 0\\
&\iff -(1-p)|\langle J_U \vert W \vert J_U \rangle|+\frac{p}{d^2}\geq 0 \quad(\Tr(W)=1)\\
&\iff p\geq 1- \frac{1}{d^2|\langle J_U \vert W \vert J_U \rangle|+1}&\\
&\Rightarrow  p^\star(U) = 1- \frac{1}{d^2|\langle J_U \vert W \vert J_U \rangle|+1}&
\end{align}

We can use the result of Casaccino \emph{et al.} to find the state that minimizes the expectation value with the witness
\begin{align}
&\underset{\rho}{\min} \left[\Tr(\rho W)\right] \text{is achieved by}\nonumber\\
&\rho=\ketbra{\nu_1}{\nu_1} \text{ where } W= \sum_k \lambda_k \ketbra{\nu_k}{\nu_k} \quad (\lambda_1\leq \lambda_2\ldots )
\end{align}
Furthermore if $\ket{\nu_1}$ is of the form $\ket{J_U}=(\I\otimes U) \sum_{j=0}^{d-1} \frac{\ket{jj}}{\sqrt{d}}$ for some $U$ then we are finished. If this is not the case (and it usually is not) then the minimal eigenvalue of $W$ at least provides an upper bound on the robustness of the optimal $U$ with respect to $W$.
\begin{align}
&U_{opt} \text{ w.r.t. } W=\underset{U\in\SU(d)}{\operatorname{argmin}} \bra{J_U}W\ket{J_U}\\
&p^\star(U_{opt}) \leq 1- \frac{1}{d^2|\lambda_1|+1}&
\end{align}
The optimal qudit ($d\geq 3$) gates that we describe were found by numerical optimization, so we cannot rule out the possibility that these gates correspond to a local minimum, although we find this unlikely. Furthermore, we cannot claim global optimality with respect to the Clifford polytope without first having a complete description of all its facets.

In the case of the qubit facets we derived in Section \ref{Example: Deriving Qubit (d=2) Clifford Polytope Facets}, the eigenvectors and eigenvalues immediately give us the globally optimal gate and its threshold noise rate. Consider the $B$-type representative facet $W_B$ of Eq.~(\ref{WBqubit}), re-scaled to have unit trace; its eigenvectors and eigenvalues are
\begin{align}
&\lambda_1=\frac{1}{2}(1-\sqrt{2}) \quad &\ket{\nu_1}=\left(
                                                      \begin{array}{c}
                                                        -\frac{1-i}{2} \\
                                                        0 \\
                                                        0 \\
                                                        \frac{1}{\sqrt{2}} \\
                                                      \end{array}
                                                    \right)\\
&\lambda_2=\frac{1}{2}(1+\sqrt{2}) \quad &\ket{\nu_2}=\left(
                                                      \begin{array}{c}
                                                        \frac{1+i}{2} \\
                                                        0 \\
                                                        0 \\
                                                        \frac{1}{\sqrt{2}} \\
                                                      \end{array}
                                                    \right)
\end{align}

\begin{align}
&\ket{\nu_1}=(I\otimes U)\sum_{j=0}^{1} \frac{\ket{jj}}{\sqrt{2}}:\quad  U=\left(
                                                                      \begin{array}{cc}
                                                                        -\e^{\frac{3\pi \i}{8}} & 0 \\
                                                                        0 & \e^{\frac{5\pi \i}{8}} \\
                                                                      \end{array}
                                                                    \right)\\
&p^\star(U)=1-\frac{1}{4\lvert \frac{1}{2}(1-\sqrt{2})\rvert+1}\approx 0.453
\end{align}
The  $W_{A^{(T)}}$-type facets have eigenvalues $\frac{1}{4}(1\pm\sqrt{3})$ which implies a threshold noise rate of at most $1-1/\sqrt{3}\approx 42\%$. Thus we have re-derived the result of Buhrman \emph{et al.} \cite{Buhrmanetal:2006} in a different way; the authors of \cite{Buhrmanetal:2006} formulated the optimization as a quadratic program and used Karush-Kuhn-Tucker conditions to ensure the solution was a global optimum.

The rest of this section is concerned with describing the most robust gates, $U_{opt}$, that we have found in dimensions $d=3,5,7$. These are analogous to the so-called $\pi/8$-gate for qubits, insofar as they are the non-Clifford unitaries that require the maximum amount of noise to enter the convex hull of Clifford gates. Intuitively it seems reasonable that such gates would take a relatively simple form, as they do below.

\subsubsection{Robust Qutrit Operation.}
The facet, $W_B^{(3)}$, used for this optimization is given explicitly in Appendix B.
\begin{align}
&\qquad\qquad U_{opt}:= \underset{U\in\SU(3)}{\operatorname{argmin}} \bra{J_U}W_B^{(3)}\ket{J_U}\\ &\bra{J_{U_{opt}}}W_B^{(3)}\ket{J_{U_{opt}}}=  \\
&\frac{1}{9}\left(3-\sqrt{3}\cos(\tfrac{\pi}{18})-6\cos(\tfrac{\pi}{9})-3\sin(\tfrac{\pi}{18})+2\sqrt{3}\sin(\tfrac{\pi}{9})\right)\nonumber\\
&\qquad\qquad U_{opt}=\left(
           \begin{array}{ccc}
             1 & 0 & 0 \\
             0 & 0 & \e^{\frac{2\pi \i}{9}} \\
             0 & \e^{-\frac{2\pi \i}{9}} & 0 \\
           \end{array}
         \right)\\
&\qquad\qquad \Rightarrow p^\star(U_{opt})\approx 78.63\%
\end{align}

\subsubsection{Robust (d=5) Qudit Operation.}
The facet, $W_B$, used for this optimization is given explicitly in Appendix B.
\begin{align}
&U_{opt}:= \underset{U\in\SU(5)}{\operatorname{argmin}} \bra{J_U}W_B\ket{J_U}\\ &\bra{J_{U_{opt}}}W_B\ket{J_{U_{opt}}}=  -\frac{4}{5}  \\
&U_{opt}=\left(
           \begin{array}{ccccc}
             1 & 0 & 0 & 0 & 0 \\
             0 & 0 & 0 & 0 & \e^{\frac{-2\pi \i}{5}} \\
             0 & 0 & 0 & \e^{\frac{4\pi \i}{5}} & 0 \\
             0 & 0 & \e^{\frac{-4\pi \i}{5}} & 0 & 0 \\
             0 & \e^{\frac{2\pi \i}{5}} & 0 & 0 & 0 \\
           \end{array}
         \right)\\
 & \Rightarrow p^\star(U_{opt})=\frac{20}{21}\approx 95.2\%
\end{align}

\subsubsection{Robust (d=7) Qudit Operation.}
The facet, $W_B$, used for this optimization is too large to reproduce but can be constructed by decoding the parity code Eq.~(\ref{paritycode}) and testing the qudit output state with respect to the phase point operator $A(0,0,0,3,2,4,2,3)$.
\begin{align}
&U_{opt}:= \underset{U\in\SU(7)}{\operatorname{argmin}} \bra{J_U}W_B\ket{J_U}\\ &\bra{J_{U_{opt}}}W_B\ket{J_{U_{opt}}}\approx -0.8411  \\
&U_{opt}=\left(
           \begin{array}{ccccccc}
             1 & 0 &0 &0 & 0 & 0 & 0 \\
             0 & 0 &0 &0 & 0 & 0 & \e^{\frac{10\pi \i}{7}} \\
             0 & 0 &0 &0 & 0 & \e^{\frac{6\pi \i}{7}} & 0 \\
             0 & 0 &0 &0 & \e^{\frac{6\pi \i}{7}} & 0 & 0 \\
             0 & 0 &0 &1 & 0 & 0 & 0 \\
             0 & 0 &\e^{\frac{6\pi \i}{7}} & 0 & 0 & 0 & 0 \\
             0 & 1 &0 &0 & 0 & 0 & 0 \\
           \end{array}
         \right)\\
 & \Rightarrow p^\star(U_{opt})\approx 97.63\%
\end{align}

\section{Summary and Open Questions.}
By using an appropriately-defined discrete Wigner function (DWF), we have found a family of qudit non-stabilizer states that are maximally robust to depolarizing noise; i.e., they are the states that are farthest outside the convex hull of all the $d$-dimensional stabilizer states. The threshold noise rate for these states takes a particularly simple form and, interestingly, these states are eigenvectors of qudit Clifford operators. Turning our attention to non-Clifford unitary gates, we found some gates which required a high rate of depolarizing before they enter the convex hull of Clifford gates (the so-called Clifford polytope). In order to find these robust gates it was first necessary to deduce facets of the Clifford polytope, and we explained a simple procedure that successfully produced many distinct facets.

An obvious open question is whether non-stabilizer qudit states, like those that we have discussed, can be purified using only stabilizer operations i.e. qudit magic state distillation. Another natural extensions of our work is to allow noise to affect stabilizer operations too, as the authors of \cite{Plenio2010} did for the qubit case. The depolarizing noise model we have used is ubiquitous in quantum information theory because of its generality and tractability. Nonetheless, other noise models are also worthy of investigation because of their relevance in fault-tolerance threshold lower bound calculations, for example, and so we highlight this as another interesting open question.

\begin{acknowledgments}
We would like to thank Ernesto Galv\~ao  and Earl Campbell for their comments regarding a previous version of this manuscript. This material is based upon work supported by the National Science
Foundation under Grant No.\ 0917244.
\end{acknowledgments}

\appendix
\section{Robust States as Eigenvectors of Clifford Operators}
Here we show that the maximally negative state
\begin{align}
\ket{\nu_1}=\frac{\ket{0}-\ket{1}}{\sqrt{2}}
\end{align}
is an eigenvector of a qudit Clifford operation for all dimensions $d$. The fact that every Clifford operation in odd dimension $d$ can be associated with a matrix $F \in SL(2,\Zd)$ in addition to a vector $\mathbf{\chi} \in \Zd^2$ results from the isomorphism
\begin{align}
\mathcal{C} \cong SL(2,\Zd) \ltimes \Zd^2,
\end{align}
established by Appleby \cite{Appleby:arxiv09}, where $\mathcal{C}$ is the Clifford group. If we specify the elements of $F$ and $\mathbf{\chi}$ as
\begin{align}
F=\left(
    \begin{array}{cc}
      \alpha & \beta \\
      \gamma & \delta \\
    \end{array}
  \right) \qquad \mathbf{\chi}=\left(
                                 \begin{array}{c}
                                   \chi_1 \\
                                   \chi_2 \\
                                 \end{array}
                               \right)
\end{align}
then Appleby provides an explicit description of the unitary matrix $C_{(F, \mathbf{\chi})}$ in terms of these elements. Initially, the Clifford operations we are interested in correspond to
\begin{align}
F=\left(
    \begin{array}{cc}
      -1 & 0\\
      -1 & -1 \\
    \end{array}
  \right) \qquad \mathbf{\chi}=\left(
                                 \begin{array}{c}
                                   0 \\
                                   0 \\
                                 \end{array}
                               \right)
\end{align}
which have matrix form
\begin{align}
C_{(F,\mathbf{\chi})}=\sum_{j=0}^{d-1} \kappa^{{j }^2} \ketbra{d-j}{j}\quad \left(\kappa=\e^{\frac{(d+1)\pi \i}{d}}\right)
\end{align}
Conjugating this Clifford operation with the Pauli operator
\begin{align}
P_{((d+1)/2,0)}= \sum_{j=0}^{d-1} \ketbra{j+(d+1)/2}{j}
\end{align}
we arrive at another Clifford operation $C^\prime$
\begin{align}
C^\prime &= P_{((d+1)/2,0)} C_{(F, \mathbf{\chi})} P_{((d+1)/2,0)} ^\dag\\
&=\sum_{j=0}^{d-1}\ket{j+(d+1)/2}\kappa^{(-j)^2}\bra{(d+1)/2-j}
\end{align}
For example, in dimension $d=7$ we have
\begin{align}
C^\prime &=\renewcommand{\arraystretch}{1}
\left( \begin{array}{ccccccc}
   \renewcommand{\arraystretch}{1.5}0 & \e^{\frac{2\pi \i}{7}} & 0 & 0 & 0 & 0 & 0 \\
   \e^{\frac{2\pi \i}{7}} & 0 & 0 & 0 & 0 & 0 & 0 \\
   0 & 0 & 0 & 0 & 0 & 0 & \e^{\frac{4\pi \i}{7}} \\
   0 & 0 & 0 & 0 & 0 & \e^{-\frac{6\pi \i}{7}} & 0 \\
    0 & 0 & 0 & 0 & 1 & 0 & 0 \\
    0 & 0 & 0 & \e^{-\frac{6\pi \i}{7}} & 0 & 0 & 0 \\
    0 & 0 & \e^{\frac{4\pi \i}{7}} & 0 & 0 & 0 & 0
 \end{array} \right)
\end{align}
Clearly $\ket{\nu_1}=\frac{\ket{0}-\ket{1}}{\sqrt{2}} $ is an eigenvector of $C^\prime$ with eigenvalue $-\e^{\frac{2\pi \i}{7}}$.
In general when $j$ takes on the values $\tfrac{d\pm 1}{2}$ in the expression for $C^\prime$ we get the following matrix entries
\begin{align*}
j=\tfrac{d-1}{2}:\qquad \ket{0}\kappa^{(\tfrac{d-1}{2})^2}\bra{1}\\
j=\tfrac{d+1}{2}:\qquad \ket{1}\kappa^{(\tfrac{d+1}{2})^2}\bra{0}
\end{align*}
Since
\begin{align}
\kappa^{(\tfrac{d-1}{2})^2}=\kappa^{(\tfrac{d+1}{2})^2}=\w^k\quad(\text{for some } k \in \Zd)
\end{align}
then $\frac{\ket{0}-\ket{1}}{\sqrt{2}}$ is an eigenvector with eigenvalue $\lambda=-\w^k$

\section{Facets of Qudit Clifford Polytope}
Here we explicitly describe the Clifford Witnesses that were used in the proving the results contained in the main text. For $d=3$ we list all distinct Witnesses (up to Clifford conjugation) that we were able to find and note whether they are true facets or not. For $d=5$ we give only the polytope facet that was relevant to $U_{opt}$ listed in the main text. The $d=7$ facet that was used to derive the threshold and optimal gate is too large to reproduce here.

\subsection{Qutrit Facets and Peaks}

\begin{align}
W_A^{(1)}=&W_A^{{u}=(0,0,0,0)}\\
W_A^{(1)}=\big(&\Pi_{(0,0|1,1)[0]}+\Pi_{(0,0|1,2)[0]}+\Pi_{(1,0|0,1)[0]}+ \\ &\Pi_{(1,0|0,2)[0]}+\Pi_{(1,0|1,1)[0]}+\Pi_{(1,0|1,2)[0]}+  \nonumber\\
&\Pi_{(1,0|2,1)[0]}+\Pi_{(1,0|2,2)[0]}-\frac{7}{3} \I\big)/3 \nonumber
\ \\\ \nonumber  \\
W_A^{(1)}: &\begin{cases}\left\vert \left\{\ket{\widetilde{J_{C_i}}} \right\} \right\vert=144\\ \operatorname{rank}\left\{\ket{\widetilde{J_{C_i}}} \right\} = 64 \text{ (True Facet) }
\end{cases}
\end{align}

\begin{align}
W_A^{(2)}=&W_A^{{u}=(0,0,1,2)}\\
W_A^{(2)}=\big(&\Pi_{(0,0|1,1)[0]}+\Pi_{(0,0|1,2)[0]}+\Pi_{(1,0|0,1)[0]}+  \\ &\Pi_{(1,0|0,2)[0]}+\Pi_{(1,0|1,1)[1]}+\Pi_{(1,0|1,2)[1]}+  \nonumber\\
&\Pi_{(1,0|2,1)[2]}+\Pi_{(1,0|2,2)[2]} -\frac{7}{3} \I\big)/3 \nonumber
\ \\\  \nonumber \\
W_A^{(2)}: &\begin{cases}\left\vert \left\{\ket{\widetilde{J_{C_i}}} \right\} \right\vert=144\\ \operatorname{rank}\left\{\ket{\widetilde{J_{C_i}}} \right\} = 62 \text{ (Not a Facet) }
\end{cases}
\end{align}

\begin{align}
W_{A^T}^{(1)}=&W_{A^T}^{{u}=(0,0,0,0)}\\
W_{A^T}^{(1)}=\big(&\Pi_{(0,0|1,1)[0]}+\Pi_{(0,0|1,2)[0]}+\Pi_{(0,1|1,0)[0]}+ \\ &\Pi_{(0,1|1,1)[0]}+\Pi_{(0,1|1,2)[0]}+\Pi_{(0,1|2,0)[0]}+  \nonumber\\
&\Pi_{(0,1|2,1)[0]}+\Pi_{(0,1|2,2)[0]} -\frac{7}{3} \I\big)/3\nonumber
\ \\\  \nonumber \\
W_{A^T}^{(1)}: &\begin{cases}\left\vert \left\{\ket{\widetilde{J_{C_i}}} \right\} \right\vert=144\\ \operatorname{rank}\left\{\ket{\widetilde{J_{C_i}}} \right\} = 64 \text{ (True Facet) }
\end{cases}
\end{align}

\begin{align}
W_{A^T}^{(2)}=&W_{A^T}^{{u}=(0,0,1,2)}\\
W_{A^T}^{(2)}=\big(&\Pi_{(0,0|1,1)[0]}+\Pi_{(0,0|1,2)[0]}+\Pi_{(0,1|1,0)[0]}+ \\ &\Pi_{(0,1|1,1)[1]}+\Pi_{(0,1|1,2)[2]}+\Pi_{(0,1|2,0)[0]}+  \nonumber\\
&\Pi_{(0,1|2,1)[1]}+\Pi_{(0,1|2,2)[2]} -\frac{7}{3} \I\big)/3 \nonumber
\ \\\ \nonumber  \\
W_{A^T}^{(2)}: &\begin{cases}\left\vert \left\{\ket{\widetilde{J_{C_i}}} \right\} \right\vert=144\\ \operatorname{rank}\left\{\ket{\widetilde{J_{C_i}}} \right\} = 62 \text{ (Not a Facet) }
\end{cases}
\end{align}

\begin{align}
W_B^{(1)}=&W_B^{{u}=(0,0,0,0)}\\
W_B^{(1)}=\big(&\Pi_{(0,0|1,1)[0]}+\Pi_{(0,0|1,2)[0]}+\Pi_{(1,2|0,0)[0]}+ \\ &\Pi_{(1,2|0,1)[0]}+\Pi_{(1,2|0,2)[0]}+\Pi_{(1,2|1,0)[0]}+  \nonumber\\ &\Pi_{(1,2|1,1)[0]}+\Pi_{(1,2|1,2)[0]}+\Pi_{(1,2|2,0)[0]}+  \nonumber\\ &\Pi_{(1,2|2,1)[0]}+\Pi_{(1,2|2,2)[0]} -\frac{10}{3} \I\big)/3 \nonumber
\ \\\ \nonumber  \\
W_B^{(1)}: &\begin{cases}\left\vert \left\{\ket{\widetilde{J_{C_i}}} \right\} \right\vert=150\\ \operatorname{rank}\left\{\ket{\widetilde{J_{C_i}}} \right\} = 64 \text{ (True Facet) }
\end{cases}
\end{align}

\begin{align}
W_B^{(2)}=&W_B^{{u}=(0,0,1,2)}\\
W_B^{(2)}=\big(&\Pi_{(0,0|1,1)[0]}+\Pi_{(0,0|1,2)[0]}+\Pi_{(1,2|0,0)[0]}+ \\ &\Pi_{(1,2|0,1)[2]}+\Pi_{(1,2|0,2)[1]}+\Pi_{(1,2|1,0)[1]}+  \nonumber\\ &\Pi_{(1,2|1,1)[0]}+\Pi_{(1,2|1,2)[2]}+\Pi_{(1,2|2,0)[2]}+  \nonumber\\ &\Pi_{(1,2|2,1)[1]}+\Pi_{(1,2|2,2)[0]} -\frac{10}{3} \I\big)/3  \nonumber
\ \\\  \nonumber \\
W_B^{(2)}: &\begin{cases}\left\vert \left\{\ket{\widetilde{J_{C_i}}} \right\} \right\vert=150\\ \operatorname{rank}\left\{\ket{\widetilde{J_{C_i}}} \right\} = 62 \text{ (Not a Facet) }
\end{cases}
\end{align}

\begin{align}
W_B^{(3)}=&W_B^{{u}=(0,2,2,0)}\\
W_B^{(3)}=\big(&\Pi_{(0,0|1,1)[0]}+\Pi_{(0,0|1,2)[0]}+\Pi_{(1,2|0,0)[2]}+  \\ &\Pi_{(1,2|0,1)[0]}+\Pi_{(1,2|0,2)[2]}+\Pi_{(1,2|1,0)[2]}+  \nonumber\\ &\Pi_{(1,2|1,1)[2]}+\Pi_{(1,2|1,2)[0]}+\Pi_{(1,2|2,0)[0]}+  \nonumber\\ &\Pi_{(1,2|2,1)[2]}+\Pi_{(1,2|2,2)[2]} -\frac{10}{3} \I\big)/3  \nonumber
\ \\\ \nonumber \\
W_B^{(3)}: &\begin{cases}\left\vert \left\{\ket{\widetilde{J_{C_i}}} \right\} \right\vert=150\\ \operatorname{rank}\left\{\ket{\widetilde{J_{C_i}}} \right\} = 64 \text{ (True Facet) }
\end{cases}
\end{align}

The witnesses that contain only $62$ linearly independent $\ket{J_{C_i}}$ are not facets, they are called peaks. The number of distinct facets (or peaks as the case may be) that can be created by conjugating with local Clifford operations is listed below where e.g. the set $\mathcal{W}_A^{(1)}$ corresponds to all facets generated by $W_A^{(1)}$.

\begin{align*}
&\lvert \mathcal{W}_A^{(1)} \rvert=864,\quad
\lvert \mathcal{W}_A^{(2)} \rvert=108\\
&\lvert \mathcal{W}_{A^T}^{(1)} \rvert=864,\quad
\lvert \mathcal{W}_{A^T}^{(2)} \rvert=108\\
&\lvert \mathcal{W}_B^{(1)} \rvert=1728,\quad
\lvert \mathcal{W}_B^{(2)} \rvert=864,\quad
\lvert \mathcal{W}_B^{(3)} \rvert=5184
\end{align*}

This gives a total of $8640$ distinct facets and $1080$ distinct peaks.

\subsection{Qudit (d=5) Facet}

The following list of $58$ 5-tuples defines the facet $W_B$ for $d=5$ qudit Clifford polytope via
\begin{align*}
W_B=\Pi_{(0,0|1,1)[0]}+\Pi_{(0,0|1,2)[0]}+\ldots
\end{align*}
.i.e. $W_B=\sum_{\text{list}} \Pi_{(x_1,x_2|z_1,z_2)[k]}$

\begin{align}
\text{list}=\left(\begin{smallmatrix}
 0 & 0 & 1 & 1 & 0 \\
 0 & 0 & 1 & 2 & 0 \\
 0 & 0 & 1 & 3 & 0 \\
 0 & 0 & 1 & 4 & 0 \\
 0 & 0 & 2 & 1 & 0 \\
 0 & 0 & 2 & 2 & 0 \\
 0 & 0 & 2 & 3 & 0 \\
 0 & 0 & 2 & 4 & 0 \\
 1 & 4 & 0 & 0 & 0 \\
 1 & 4 & 0 & 1 & 1 \\
 1 & 4 & 0 & 2 & 3 \\
 1 & 4 & 0 & 3 & 1 \\
 1 & 4 & 0 & 4 & 0 \\
 1 & 4 & 1 & 0 & 0 \\
 1 & 4 & 1 & 1 & 0 \\
 1 & 4 & 1 & 2 & 1 \\
 1 & 4 & 1 & 3 & 3 \\
 1 & 4 & 1 & 4 & 1 \\
 1 & 4 & 2 & 0 & 1 \\
 1 & 4 & 2 & 1 & 0
\end{smallmatrix}\right)\left(\begin{smallmatrix}
 1 & 4 & 2 & 2 & 0 \\
 1 & 4 & 2 & 3 & 1 \\
 1 & 4 & 2 & 4 & 3 \\
 1 & 4 & 3 & 0 & 3 \\
 1 & 4 & 3 & 1 & 1 \\
 1 & 4 & 3 & 2 & 0 \\
 1 & 4 & 3 & 3 & 0 \\
 1 & 4 & 3 & 4 & 1 \\
 1 & 4 & 4 & 0 & 1 \\
 1 & 4 & 4 & 1 & 3 \\
 1 & 4 & 4 & 2 & 1 \\
 1 & 4 & 4 & 3 & 0 \\
 1 & 4 & 4 & 4 & 0 \\
 2 & 3 & 0 & 0 & 0 \\
 2 & 3 & 0 & 1 & 0 \\
 2 & 3 & 0 & 2 & 3 \\
 2 & 3 & 0 & 3 & 4 \\
 2 & 3 & 0 & 4 & 3 \\
 2 & 3 & 1 & 0 & 3 \\
 2 & 3 & 1 & 1 & 0
\end{smallmatrix}\right)\left(\begin{smallmatrix}
 2 & 3 & 1 & 2 & 0 \\
 2 & 3 & 1 & 3 & 3 \\
 2 & 3 & 1 & 4 & 4 \\
 2 & 3 & 2 & 0 & 4 \\
 2 & 3 & 2 & 1 & 3 \\
 2 & 3 & 2 & 2 & 0 \\
 2 & 3 & 2 & 3 & 0 \\
 2 & 3 & 2 & 4 & 3 \\
 2 & 3 & 3 & 0 & 3 \\
 2 & 3 & 3 & 1 & 4 \\
 2 & 3 & 3 & 2 & 3 \\
 2 & 3 & 3 & 3 & 0 \\
 2 & 3 & 3 & 4 & 0 \\
 2 & 3 & 4 & 0 & 0 \\
 2 & 3 & 4 & 1 & 3 \\
 2 & 3 & 4 & 2 & 4 \\
 2 & 3 & 4 & 3 & 3 \\
 2 & 3 & 4 & 4 & 0
\end{smallmatrix}\right)
\end{align}

Note that
\begin{align}
W_B: &\begin{cases}\left\vert \left\{\ket{\widetilde{J_{C_i}}} \right\} \right\vert=2420\\
\operatorname{rank}\left\{\ket{\widetilde{J_{C_i}}} \right\} = (d^2-1)^2=576\quad(\text{True Facet})
\end{cases}
\end{align}


\begin{thebibliography}{99}
\bibitem{JozsaLinden2003}
R.~Jozsa and N.~Linden,
``On the role of entanglement in quantum-computational speed-up'',
\emph{Proceedings of the Royal Society A: Mathematical, Physical and Engineering Sciences} Volume 459, number 2036, pp.~2011--2032,  (2003).

\bibitem{Eastin:arxiv10}
B.~Eastin,
 ``Simulating Concordant Computations'',
arXiv:quant-ph/1006.4402, (2010).

\bibitem{Raussendorf:arxiv09}
R.~Raussendorf,
 ``Quantum computation, discreteness, and contextuality'',
arXiv:quant-ph/0907.5449, (2009).

\bibitem{NielsenChuang:2000}
M.~A.~Nielsen and I.~L.~Chuang,
\newblock {\em Quantum Computation and Quantum Information}.
\newblock Cambridge University Press, Cambridge, (2000).

\bibitem{Buhrmanetal:2006}
 H.~Buhrman, R.~Cleve, M.~Laurent, N.~Linden, A.~Schrijver and F.~Unger,
``New Limits on Fault-Tolerant Quantum Computation'',
 \emph{Annual IEEE Symposium on Foundations of Computer Science,}
 pp.~411--419, (2006).


 \bibitem{Plenio2010}
M.~B.~Plenio and S.~Virmani,
``Upper bounds on fault tolerance thresholds of noisy Clifford-based quantum computers''
\newblock New Journal of Physics \textbf{12}, number 3, 033012 , (2010).


\bibitem{Knill05}
E.~Knill ,
``Quantum Computing with Realistically Noisy Devices"
\newblock Nature \textbf{434}, pp.~39--44, (2005).

\bibitem{AliferisCross07}
P.~Aliferis and A.~W.~Cross,
``Subsystem Fault Tolerance with the Bacon-Shor Code"
\newblock Phys.~Rev.~Lett. \textbf{98}, 220502, (2007).

\bibitem{AliferisPreskill08}
P.~Aliferis and J.~Preskill,
``Fault-tolerant quantum computation against biased noise"
\newblock Phys.~Rev.~A. \textbf{78}, 052331, (2008).

\bibitem{ReichardtThresh09}
B.~Reichardt,
``Error-Detection-Based Quantum Fault-Tolerance Threshold"
\newblock Algorithmica \textbf{55}, issue 3, pp.~517--556 (2009).

\bibitem{FujiiYamamoto10}
K.~Fujii and K.~Yamamoto,
``Topological one-way quantum computation on verified logical cluster states"
\newblock Phys.~Rev.~A. \textbf{82}, 060301, (2010).

\bibitem{Wang10}
D.~S.~Wang, A.~G.~Fowler and L.~C.~L.~Hollenberg,
``Surface code quantum computing with error rates over 1$\%$"
\newblock Phys.~Rev.~A. \textbf{83}, 020302, (2011).


\bibitem{ReichardtMagic09}
B.~Reichardt,
``Quantum universality by state distillation "
\newblock Quantum Inf. Comput. \textbf{9},  pp.~1030--1052 (2009).

\bibitem{FreedmanNayakWalker:2006}
M.~Freedman, C.~Nayak and K.~Walker,
``Towards universal topological quantum computation in the $\nu = (5/2)$ fractional quantum Hall state'',
\newblock Phys.~Rev.~B \textbf{73}, 245307 (2006).

\bibitem{Georgiev:2006}
L.~S.~Georgiev,
``Topologically protected gates for quantum computation with
  non-abelian anyons in the Pfaffian quantum Hall state'',
\newblock Phys.~Rev.~B \textbf{74}, 235112 (2006)''

\bibitem{BravyiKitaev:2005}
S.~Bravyi and A.~Kitaev,
`` Universal quantum computation with ideal Clifford gates and noisy ancillas'',
\newblock Phys.~Rev.~A \textbf{71}, 022316 (2005).


\bibitem{Reichardt:2005}
Ben~W.~Reichardt,
`` Quantum universality from magic states distillation applied to CSS codes'',
\newblock Quant.~Inf.~Proc., \textbf{4}(3):251--264 (2005).

\bibitem{CampbellBrowne:2010}
E.~T.~Campbell and D.~E.~Browne,
``Bound States for Magic State Distillation in Fault-Tolerant Quantum Computation"
\newblock Phys.~Rev.~Lett. \textbf{104}, 030503, (2010).


\bibitem{CampbellBrowne:2009}
E.~T.~Campbell and D.~E.~Browne,
``On the Structure of Protocols for Magic State Distillation'',
in \emph{Theory of Quantum Computation, Communication, and Cryptography}, (editor: Childs, Andrew and Mosca, Michele),
\emph{Lecture Notes in Computer Science,} Volume 5906, (Springer Berlin / Heidelberg, 2009), pp.~20--32.

\bibitem{Campbell:arxiv10}
E.~T.~Campbell,
 ``Catalysis and activation of magic states in fault tolerant architectures'',
arXiv:1010.0104, (2010).

 \bibitem{WvDMH:2009}
W.~van~Dam and M.~Howard,
``Tight Noise Thresholds for Quantum Computation with Perfect Stabilizer Operations"
\newblock Phys.~Rev.~Lett. \textbf{103}, 170504, (2009).


 \bibitem{Galvao:2005}
 E.~F.~Galv\~ao,
``Discrete Wigner functions and quantum computational speedup''
\newblock Phys.~Rev.~A. \textbf{71}, 042302, (2005).


 \bibitem{Cormick:2006}
 C.~Cormick,  E.~F.~Galv\~ao, D.~Gottesman, J.~Pablo~Paz, and Arthur~O.~Pittenger ,
``Classicality in discrete Wigner functions"
\newblock Phys.~Rev.~A. \textbf{73}, 012301, (2006).

 \bibitem{Gross:2006}
 D.~Gross,
``Hudson's theorem for finite-dimensional quantum systems''
\newblock J.~Math.~Phys. \textbf{47}, number 12, 122107, (2006).

 \bibitem{Wootters1987}
W.~K.~Wootters,
``A Wigner-function formulation of finite-state quantum mechanics''
\newblock Annals of Physics \textbf{176}, number 1, pp.~1--21 , (1987).


 \bibitem{Gibbons:2004}
K.~S.~Gibbons, M.~J.~Hoffman, and  W.~K.~Wootters,
``Discrete phase space based on finite fields''
\newblock Phys.~Rev.~A. \textbf{70}, 062101, (2004).


\bibitem{Casaccino:2008}
A.~Casaccino,  E.~F.~Galv\~ao,  and S.~Severini,
``Extrema of discrete Wigner functions and applications"
\newblock Phys.~Rev.~A. \textbf{78}, 022310, (2008).

\bibitem{appleby:012102}
 D.~M.~Appleby, I.~Bengtsson and S.~Chaturvedi,
 ``Spectra of phase point operators in odd prime dimensions and the extended Clifford group'',
 \emph{Journal of Mathematical Physics,}
 Volume 49, number 1, pp.~012102, (2008).


\bibitem{Gottesman1999}
 D.~Gottesman,
``Fault-Tolerant Quantum Computation with Higher-Dimensional Systems'',
in \emph{Quantum Computing and Quantum Communications}, (editor: Colin Williams),
\emph{Lecture Notes in Computer Science,} Volume 1509, (Springer Berlin / Heidelberg, 1999), pp.~302--313.

\bibitem{Beigi:arxiv06}
S.~Beigi, M.~Bahramgiri,
 ``Graph States Under the Action of Local Clifford Group in Non-Binary Case'',
arXiv:quant-ph/0610267v2, (2006).



\bibitem{Chazelle1991}
 B.~Chazelle,
 ``An optimal convex hull algorithm and new results on cuttings'',
 \emph{Proceedings of Thirty-
Second Annual IEEE Symposium on Foundations of Computer Science (FOCS)}
(1991), pp.~29--38.


 \bibitem{Avis2000Revised}
D.~Avis,
``A revised implementation of the reverse search vertex enumeration algorithm''
in \emph{Polytopes - Combinatorics and Computation (Oberwolfach Seminars)} , (editors: Kalai, G. and Ziegler, G.), pp.~177--198, Birkh\"{a}user Basel, (2000).

\bibitem{Fukuda1996}
 K.~Fukuda, and A.~Prodon,
``Double description method revisited''
in \emph{Combinatorics and Computer Science} , (editors: Deza, Michel and Euler, Reinhardt and Manoussakis, Ioannis),\emph{Lecture Notes in Computer Science,}  volume 1120, (Springer Berlin / Heidelberg 1996) pp.~91--111.



\bibitem{Bagan:2003}
 E.~Bagan, M.~Baig,  and R.~Mu\~noz-Tapia,
``Minimal measurements of the gate fidelity of a qudit map",
\newblock Phys.~Rev.~A. \textbf{67}, 014303, (2003).


 \bibitem{Masanes2003}
Ll.~Masanes,
``Tight Bell inequality for d-outcome measurements correlations.''
\newblock Quantum Inf. Comput. \textbf{3}, pp.~345--358 , (2003).

\bibitem{Bandyopadhyay2008}
S.~Bandyopadhyay, P.~O.~Boykin, V.~Roychowdhury, and F.~Vatan,
 ``A New Proof for the Existence of Mutually Unbiased Bases'',
 \emph{Algorithmica,}
 Volume 34, issue 4, pp.~512--528, (2002).


\bibitem{Appleby:arxiv09}
 D.~M.~Appleby,
 ``Properties of the extended Clifford group with applications to SIC-POVMs and MUBs'',
arXiv:quant-ph/0909.5233, (2009).

\end{thebibliography}
\end{document}